\documentclass[a4paper,11pt]{article}
\usepackage{pos}

\usepackage{amsmath,amsfonts,dcolumn}
\usepackage{graphicx}
\usepackage{subcaption}

\newcommand{\Lat}[1]{\textit{#1}} % Latin phrases in italic (e.g., i.e. etc.)
\newcommand{\M}{\mathcal{M}}

\newcommand{\F}{\mathcal{F}}

\renewcommand{\Im}{\text{Im}\,}
\renewcommand{\Re}{\text{Re}\,}

\newcommand{\iu}{\mathrm{i}}

\newcommand{\dff}{\mathrm{d}}

\newcommand{\BW}{\text{BW}}

\newcommand{\Born}{\text{Born}}

\newcommand{\elm}{\text{em}}

\newcommand{\disc}{\text{disc}}
\newcommand{\MO}{\text{MO}}
\newcommand{\exch}{\text{exch}}

\usepackage[dvipsnames]{xcolor}

\title{Towards an improved understanding of $a_1$ and $a_2$ transition form factors}
\author*[a]{Hannah Schäfer}

\affiliation{Albert Einstein Center for Fundamental Physics,\\ Institute for Theoretical Physics, University of Bern, Sidlerstrasse 5, 3012 Bern, Switzerland}

\emailAdd{hannah.schaefer@unibe.ch}

\abstract{The isovector axial-vector and tensor meson states $a_1$ and $a_2$ have garnered interest in the last years since they appear as intermediate states in the hadronic light-by-light (HLbL) scattering contribution to $(g-2)_\mu$. 
Their dynamical behaviour including their transition form factors is poorly understood to this day, which also affects the uncertainty budget of the HLbL contribution.
In order to improve the situation, we construct the $a_1$ and $a_2$ as resonances in the $\rho \pi$ system, which allows us to express the relevant amplitudes in terms of known form factors and properties of the resonances.
With this, we aim to obtain a phenomenologically well motivated parameterisation of the transition form factors built from the general principles of analyticity, unitarity, and crossing symmetry.}

\FullConference{14th International Workshop on e$^+$e$^-$ collisions from Phi to Psi 2026 (PHIPSI26)\\
8-11 June 2026\\
Pisa, Italy\\}

\begin{document}

%These lines have to go after \begin{document} and adjust the referenced labels for \autoref.
%---------------------------------------------------------------------------------------------------
\renewcommand{\figureautorefname}{Fig.}
\renewcommand{\tableautorefname}{Table}
\renewcommand{\chapterautorefname}{Ch.}
\renewcommand{\sectionautorefname}{Sec.}
\renewcommand{\subsectionautorefname}{Sec.}
\renewcommand{\appendixautorefname}{App.}
\def\equationautorefname~#1\null{Eq.~(#1)\null}
%---------------------------------------------------------------------------------------------------

\maketitle

%---------------------------------------------------------------------------------------------------
\section{Introduction}
\label{sec:intro}
\begin{figure}[t]
    \centering
    \includegraphics[]{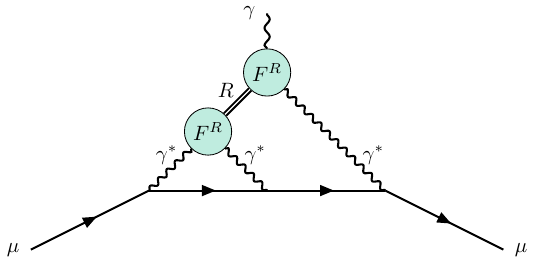}
    \caption{Contribution of a single-resonance state to HLbL, with TFFs $F^R \equiv F^R(q_1^2,q_2^2)$.}
    \label{fig:HLbL_res}
\end{figure}

%---------------------------------------------------------------------------------------------------
The anomalous magnetic moment of the muon $(g-2)_\mu = 2 a_\mu$ has been examined for a discrepancy between the experimental result~\cite{Muong-2:2023cdq,Muong-2:2006rrc,Muong-2:2026qnz} %Muong-2:2021ojo,
and the Standard Model (SM) calculation~\cite{Aliberti:2025beg} for a long time, with the hadronic contributions adding most to the uncertainty budget on the theory side. 
They consist of the hadronic vacuum polarisation (HVP), which is currently being scrutinised, and hadronic light-by-light scattering (HLbL), where huge progress has been made within the last years by means of an improved dispersive framework~\cite{Hoferichter:2024fsj}, a better description of dominant hadronic contributions (\Lat{e.g.}, Refs.~\cite{Hoferichter:2023tgp,Holz:2024diw}), and a better understanding of and control over the short-distance and mixed regions and how they can be matched to the non-perturbative region~\cite{Bijnens:2024jgh,Bijnens:2022itw,Bijnens:2021jqo,Hoferichter:2024vbu}.
While the lowest-lying hadronic states in the HLbL sector, the pseudoscalar poles, $\pi/K$ boxes, and $S$-wave rescattering contributions, are well understood by now, there are still gaps in our knowledge of intermediate states of higher spin that are reflected by a larger uncertainty attributed to their contribution to $a_\mu$ and to the matching to short-distance constraints.
The description of discrete hadronic intermediate states in the dispersive framework requires knowledge about their transition form factors (TFFs) for space-like photon virtualities, where either one or both photons are off-shell; see \autoref{fig:HLbL_res}.
For the triplet of axial-vector mesons, the $f_1(1285)$ TFFs have been analysed in a vector-meson-dominance (VMD)-motivated framework, taking into account all available experimental data at the time~\cite{Zanke:2021wiq,Hoferichter:2023tgp}, while the behaviour of the heavier $f_1^\prime(1420)$ and the isovector $a_1(1320)$ is hard to fix from data. 
Their TFFs are therefore estimated from $\mathrm{U}(3)$ symmetry.
For the $a_1$, there is additionally a dispersive description for $F_2^{a_1}(q_1^2,0)$ in the framework of the $VVA$ correlator~\cite{Ludtke:2024ase}, which matches the $\mathrm{U}(3)$ result rather well.
In the case of the tensor meson triplet, the estimate of the 2025 Theory Initiative White Paper (WP25)~\cite{Aliberti:2025beg} is based on a simple quark model for $F_1^{T}(q_1^2, q_2^2)$, while the other four TFFs vanish.
There is ongoing work on the tensor TFFs in holographic QCD~\cite{Cappiello:2025fyf} and in resonance chiral theory (RChT)~\cite{Estrada:2025bty}, and specifically on the $f_2(1270)$ also in dispersive frameworks~\cite{Zillinger:2026ulk,Kaziukenas:2026Pisa}.

While the isoscalar $f_2$ occurs in the $D$-wave of the $\pi\pi$ system, which has been understood rather well by now~\cite{Hoferichter:2019nlq}, the isovector $a_2$ is associated with a $3\pi$ system, as well as the $a_1$, mostly coupling to the $\rho\pi$ channel of that system~\cite{ParticleDataGroup:2026}.
Consequently, understanding the coupling of $\rho\pi$ to two virtual photons and identifying the contributions of the hadronic resonances of interest is a way towards describing the $a_1$ and $a_2$ TFFs.
We outline here some steps towards this goal; earlier work on this has already been presented in Ref.~\cite{Schafer:2025gfa}.
It has, however, become clear that the extraction of TFFs from a direct loop calculation as demonstrated there would require further assumptions.
In the following, we explain a different strategy where the TFFs are related to a Muskhelishvili-Omnès solution of the $\rho\pi \to \gamma^*\gamma^*$ system.
%---------------------------------------------------------------------------------------------------
\section{The \boldmath{$\rho\pi\to\gamma^*\gamma^*$} system}
\label{sec:RhoPiGG}
%---------------------------------------------------------------------------------------------------
%
\begin{figure}
	\centering
	\includegraphics[scale=1]{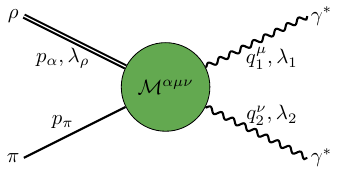}
	\caption{The process $\rho\pi \to \gamma^*\gamma^*$ with definitions of momenta and helicities.}
	\label{fig:ggPiRhoBlob}
\end{figure}
The matrix element
\begin{align}
    &\iu \int \dff^4 z e^{-\iu q_1 \cdot z} \langle 0 \vert T\{j_{\elm}^\mu(z) j_{\elm}^\nu(0)\} \vert \rho(p) \pi(p_\pi) \rangle 
    = \epsilon_{\alpha}(p,\lambda_\rho) \M_{\rho\pi\gamma^*\gamma^*}^{\mu\nu\alpha}(q_1,q_2,p)
\end{align}
of the amplitude
\begin{align}
&\langle  \gamma^*(q_1,\lambda_1) \gamma^*(q_2,\lambda_2) \vert \rho(p) \pi(p_\pi) \rangle =  \notag \\
& -e^2 (2\pi)^4 \delta(q_1+q_2-p -p_\pi)
\epsilon_{\mu}^*(q_1,\lambda_1) \epsilon_{\nu}^*(q_2,\lambda_2) \hspace{-0.1cm} \int\hspace{-0.1cm} \dff^4 z\, e^{-\iu q_1 \cdot z} \langle 0 \vert T\{j_{\elm}^\mu(z) j_{\elm}^\nu(0)\} \vert \rho(p) \pi(p_\pi)  \rangle \notag \\
&= \iu e^2 (2\pi)^4 \delta(q_1+q_2-p -p_\pi) \epsilon_{\mu}^*(q_1,\lambda_1) \epsilon_{\nu}^*(q_2,\lambda_2) \epsilon_{\alpha}(p,\lambda_\rho) \M_{\rho\pi\gamma^*\gamma^*}^{\mu\nu\alpha}(q_1,q_2,p)
\end{align}
can be decomposed into tensor structures and scalar functions as
\begin{align}
	\M_{\rho\pi\gamma^*\gamma^*}^{\mu\nu\alpha}(q_1,q_2,p) = \sum_{i=1}^{13} \F_i(s,t,u; q_1^2,q_2^2) T_i^{\mu\nu\alpha}(q_1,q_2,p).
\end{align}
Here, we treat the vector and the pseudoscalar particle as asymptotic states and switch back and forth between treating the photons as external states and as currents using the LSZ formula; see \autoref{fig:ggPiRhoBlob} for the definition of momenta and helicities. 
The construction of a gauge-invariant basis of tensor structures $T_i^{\mu\nu\alpha}$ and scalar functions $\mathcal{F}_i$ for $\mathcal{M}_{\rho\pi\gamma^*\gamma^*}^{\mu\nu\alpha}$ via a Bardeen--Tung--Tarrach (BTT) decomposition~\cite{Bardeen:1968ebo,Tarrach:1975tu} has been presented in Ref.~\cite{Schafer:2025gfa}, and there it is also explained how kinematic zeros and singularities are removed and redundancies in the basis are identified.

\begin{figure}[t]
    \centering
    \subfloat[]
    {\includegraphics[]{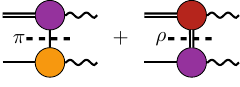}}
    \qquad
    \subfloat[]
    {\includegraphics[]{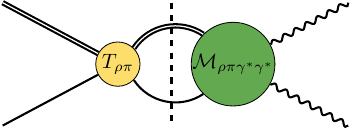}}
    \caption{Cuts for the Born terms of $\rho\pi \to \gamma^*\gamma^*$ (a) and $\rho\pi$ rescattering (b). 
    The coloured blobs correspond to the pion vector form factor (orange), the $\rho$ form factors (red), and the $\rho\pi$ TFF (purple), respectively, and the amplitudes $\rho\pi \to \rho\pi$ (yellow) and $\rho\pi \to \gamma^*\gamma^*$ (green).}
    \label{fig:BornTerms_RhoPiCut}
\end{figure}

Since the scalar functions $\F_i(s,t,u; q_1^2, q_2^2)$ contain only dynamical singularities, one can set up dispersion relations for them,
\begin{align}\label{eq:DR_SF_BornRHC}
	\F_i(s, t, u; q_1^2, q_2^2) = \F_i^{\Born}(s, t, u; q_1^2, q_2^2) + \frac{1}{2\pi \iu} \int_{(M_\rho+M_\pi)^2}^\infty \dff s^\prime \, \frac{\disc_{s^\prime}^{\rho\pi} \F_i(s^\prime, t^\prime, u^\prime; q_1^{2}, q_2^{2})}{s^\prime - s},
\end{align}
\Lat{e.g.}, in terms of hyperbolic dispersive variables $t^\prime$, $u^\prime$~\cite{Hite:1973pm}. 
The Born terms are in this case given by exchanges of $\rho$ and $\pi$ in the $t$- and $u$-channel, which can be reconstructed dispersively using knowledge about the pion vector form factor, the $\rho$ form factors, and the $\rho\pi$ TFF; see the left-hand side of \autoref{fig:BornTerms_RhoPiCut}.
The dispersive part contains the discontinuity with regard to a $\rho\pi$ intermediate state, which is associated with rescattering in this channel.
On the right-hand side of \autoref{fig:BornTerms_RhoPiCut}, this is depicted by the rescattering matrix element $T_{\rho\pi} \equiv T_{\rho\pi}^{\alpha\beta}(p,p^\prime)$.
\autoref{eq:DR_SF_BornRHC} can be generalised by the inclusion of additional integrals over LHCs; see, \Lat{e.g.}, Refs.~\cite{Moussallam:2013una,Hoferichter:2019nlq}.

An equation of the type of \autoref{eq:DR_SF_BornRHC} is solved by a Muskhelishvili-Omnès (MO) solution~\cite{Muskhelishvili:1942,Omnes:1958hv}, which sums up all rescattering effects and takes into account Born terms and possibly left-hand cuts (LHCs). 
Such a solution is, however, constructed for scalar quantities depending explicitly on one variable, while the scalar functions here depend on two independent Mandelstam variables (which can be rewritten to a dependence on $s$ and the scattering angle $z_s$). 
We therefore need to consider helicity amplitudes
\begin{align}
    e^{\iu \varphi (\lambda_\rho - \lambda_1 + \lambda_2)} H_{\rho\pi\gamma^*\gamma^*}^{\lambda_1\lambda_2;\lambda_\rho}(s, z_s) = \epsilon_{\alpha}(p, \lambda_\rho) \epsilon_{\mu}^*(q_1,\lambda_1) \epsilon_{\nu}^*(q_2,\lambda_2) \, \M_{\rho\pi\gamma^*\gamma^*}^{\mu\nu\alpha}(q_1,q_2,p)
\end{align}
and 
\begin{align}
    e^{\iu \varphi(\lambda_\rho^\prime - \lambda_\rho)} H_{\rho\pi}^{\lambda_\rho \lambda_\rho^\prime}(s, z_s) &= \epsilon_\alpha(p,\lambda_\rho) \epsilon_\beta^*(p^\prime,\lambda_\rho^\prime) \, T_{\rho\pi}^{\alpha\beta}(p, p^\prime) ,
\end{align}
and subsequently project them to partial waves $m_{\lambda_1 \lambda_2;\lambda_\rho}^J(s)$ and $t_{\lambda_\rho\lambda_\rho^\prime}^J(s)$ depending only on one variable, while the angular dependence resides in the Wigner-$d$ functions; $J$ denotes the total angular momentum.
This yields a system of coupled equations for the $m_{\lambda_1 \lambda_2;\lambda_\rho}^J(s) =: m_j^J(s)$ since each helicity amplitude can depend on all scalar functions,
\begin{align}\label{eq:SFtoHA}
	H_{\rho\pi\gamma^*\gamma^*}^{\lambda_1\lambda_2;\lambda_\rho}(s, z_s) \equiv H_{\rho\pi\gamma^*\gamma^*}^j(s, z_s) = \sum_{i=1}^{13} c_{ji}(s,z_s) \F_i(s,t,u; q_1^2, q_2^2) ,
\end{align}
with $z_s(s,t,u)$, and can additionally contribute to different partial waves,
\begin{align}\label{eq:PWE_H}
	H_{\rho\pi\gamma^*\gamma^*}^{\lambda_1 \lambda_2; \lambda_\rho}(s, z_s) = \sum_{J=0}^\infty (2J+1) d_{(\lambda_1-\lambda_2) \lambda_\rho}^J (z_s) m_{\lambda_1 \lambda_2; \lambda_\rho}^{J}(s) .
\end{align}
The dispersion relations for the partial-wave-projected helicity amplitudes are then given by
\begin{align}
    m_j^J(s) = m_j^{\Born,J}(s) + \sum_{J^\prime, j^\prime} \frac{1}{2 \pi \iu} \int_{(M_\rho+M_\pi)^2}^\infty \dff s^\prime K_{J J^\prime}^{j j^\prime}(s, s^\prime) \disc_{s^\prime}^{\rho\pi} m_{j^\prime}^{J^\prime}(s^\prime),
\end{align}
with kernel functions $K_{J J^\prime}^{j j^\prime}(s, s^\prime)$ emerging from the integration kernels in \autoref{eq:DR_SF_BornRHC}, the coefficients in \autoref{eq:SFtoHA}, and the Wigner-$d$ functions via an integration over the angle $z_s$. 

%---------------------------------------------------------------------------------------------------
\section[$\rho\pi$ rescattering]{\boldmath{$\rho\pi$} rescattering}
\label{sec:RhoPiRhoPi}
%---------------------------------------------------------------------------------------------------
Unitarity determines the discontinuity to be
\begin{align}
    \disc_s^{\rho\pi} m_{\lambda_1 \lambda_2; \lambda_\rho}^J (s) \sim \sum_{\lambda_\rho^\prime} t_{\lambda_\rho^\prime\lambda_\rho}^{J*}(s) m_{\lambda_1 \lambda_2; \lambda_\rho^\prime}^J (s).
\end{align}
Hence one needs to constrain $\rho\pi$ rescattering in order to construct the MO solution.
Watson's theorem implies that the imaginary part of an amplitude is related to the rescattering phase shift~\cite{Watson:1954uc}. 
One way to approach the phase shift is to relate it again to the resonances in which we are interested since for a given partial wave, the respective resonance is assumed to dominate the phase movement in the region of interest. 
For this, we calculate amplitudes for the process $\rho\pi \to \rho\pi$ via an intermediate one-particle resonance state $M$.
A similar machinery as for $\rho\pi \to \gamma^*\gamma^*$ can be employed for this amplitude, \Lat{i.e.}, finding a basis of tensor structures and scalar functions which respect all symmetries (in this case transversality of the $\rho$ field instead of gauge invariance), calculation of helicity amplitudes, mapping of the amplitude with intermediate state $M$ to the helicity basis, and partial-wave projection.
The latter serves here as a cross-check since a $J^{PC}$ resonance state should only contribute pole terms to the $J$-th partial wave and additionally show a specific pattern in the helicity amplitudes due to parity.
The basis consists in this case of 5 tensor structures, corresponding to 5 independent helicity amplitudes
\begin{align}\label{eq:HelAmpsRhoPiRhoPi}
	H_{\rho\pi}^{++} &= H_{\rho\pi}^{1} = H_{\rho\pi}^{--}, \qquad &H_{\rho\pi}^{+-} &= H_{\rho\pi}^{2} = H_{\rho\pi}^{-+}, &\notag\\
    H_{\rho\pi}^{+0} &= H_{\rho\pi}^{3} = -H_{\rho\pi}^{-0}, \qquad &H_{\rho\pi}^{0+} &= H_{\rho\pi}^{4} = -H_{\rho\pi}^{0-}, \qquad &H_{\rho\pi}^{00} &= H_{\rho\pi}^{5} ,
\end{align}
which can be projected to partial waves via
\begin{align}
    t_{\lambda_\rho\prime \lambda_\rho}^J(s) = \frac{1}{2} \int_{-1}^1 \dff z_s\, d_{\lambda_\rho \lambda_\rho^\prime}^J (z_s) H_{\rho\pi}^{\lambda_\rho \lambda_\rho^\prime} (s, z_s).
\end{align}
In order to understand the tensor and helicity structure, it is enough to write down the amplitude for a stable intermediate state $M$ coupling to $\rho\pi$ via a tensor structure $\M_{M\rho\pi}^{\mu\tilde{\alpha}}$, where $\mu$ gets contracted with the $\rho$ polarisation and $\tilde{\alpha}$ with the polarisation vector of $M$ (which might have any spin, so $\tilde{\alpha}$ is potentially a multi-index). 
This simplified amplitude can be expressed in terms of a dispersion relation as
\begin{align}\label{eq:realpole_rhopirhopi}
    T_{\rho\pi, M}^{\mu\nu} = \frac{1}{2\pi\iu}\int \dff s^\prime \frac{\disc_{s^\prime}^M T_{\rho\pi}^{\mu\nu}}{s^\prime - s} = \int \dff s^\prime \frac{\delta(s^\prime-M_{M}^2) \M_{M \rho\pi}^{\mu\tilde{\alpha}*} \M_{M \rho\pi}^{\nu\tilde{\beta}} s_{\tilde{\alpha}\tilde{\beta}}^M}{s^\prime - s}  = \frac{\M_{M \rho\pi}^{\mu\tilde{\alpha}*} \M_{M \rho\pi}^{\nu\tilde{\beta}} s_{\tilde{\alpha}\tilde{\beta}}^M}{M_M^2 -s},
\end{align}
where $s_{\tilde{\alpha}\tilde{\beta}}^M$ denotes the polarisation sum of $M$.
This procedure has been performed for the $a_1$ and $a_2$, and additionally, in order to understand the partial-wave and parity patterns, for the $\pi(1300) \equiv \pi_0$ (the $a_0(980)$ is forbidden by parity), the $\pi_1(1600)$, and the $\pi_2(1670)$.
We find the following pattern for the tuples of helicity amplitudes in the ordering of \autoref{eq:HelAmpsRhoPiRhoPi},
\begin{align} \label{eq:PWsRhoPiRhoPi}
    \left(t_{\pi_0}^0\right) &= \frac{}{} \left( 0,0,0,0, -\frac{C_{\pi_0\rho\pi}^2 \lambda_{\rho\pi}(s)}{M_\rho^2(s-M_{\pi_0}^2)} \right), \notag\\
    \left(t_{a_1}^1\right) &= \frac{C_{a_1\rho\pi}^2}{12 (s- M_{a_1}^2)} \Bigg( -(s+ M_\rho^2 -M_\pi^2)^2, -(s+ M_\rho^2 - M_\pi^2)^2, \sqrt{2s} M_\rho (s+ M_\rho^2 -M_\pi^2),\notag\\
    &\hspace{2.7cm} \sqrt{2s} M_\rho(s+ M_\rho^2 -M_\pi^2), 4 M_\rho^2 s \Bigg), \notag\\
    \left(t_{\pi_1}^1\right) &=  \left( -\frac{C_{\pi_1\rho\pi}^2 \lambda_{\rho\pi}(s)}{12 (s- M_{\pi_1}^2)}, \frac{C_{\pi_1\rho\pi}^2 \lambda_{\rho\pi}(s)}{12 (s- M_{\pi_1}^2)}, 0, 0, 0 \right), \notag\\
    \left(t_{a_2}^2\right) &= \left( -\frac{C_{a_2\rho\pi}^2 \lambda_{\rho\pi}^2(s)}{40 s (s- M_{a_2}^2)}, \frac{C_{a_2\rho\pi}^2 \lambda_{\rho\pi}^2(s)}{40 s (s- M_{a_2}^2)}, 0, 0, 0 \right), \notag\\
    \left(t_{\pi_2}^2\right) &= \frac{C_{\pi_2 \rho\pi}^2 \lambda_{\rho\pi}(s)}{40s(s-M_{\pi_2}^2)} \Bigg( -(s+M_\rho^2-M_\pi^2)^2, -(s+M_\rho^2-M_\pi^2)^2 , \notag\\
    &\hspace{2.7cm}-\frac{4 M_\rho \sqrt{s}}{\sqrt{3}}(s+M_\rho^2-M_\pi^2) , \frac{4 M_\rho \sqrt{s}}{\sqrt{3}}(s+M_\rho^2-M_\pi^2) , \frac{16 M_\rho^2 s}{3} \Bigg). 
\end{align}
The partial waves not quoted explicitly either vanish or, in the case of the $P$-waves for the tensor mesons, do not have resonance poles.
From these amplitudes it can be seen that there is one phase per resonance which contributes to the different helicity amplitudes with different kinematic prefactors related to the absolute of the three-momentum and the energy of the $\rho$.

In order to obtain a phase shift, one needs to replace the real pole in $T_{\rho\pi, M}^{\mu\nu}$ by a complex pole position.
The phase shift associated with this resonance is then given by 
\begin{align}
    \delta_{M}(s) = \arctan\left(\frac{\Im t_{M,j}(s)}{\Re t_{M,j}(s)}\right) ,
\end{align}
where $t_{M,j}(s)$ denotes the $j$-th partial-wave-projected helicity amplitude; with a simple Breit--Wigner parameterisation, this is given by
\begin{align}
    \delta_{M}^\BW(s) = \arctan\left( \frac{M_M \Gamma_M(s)}{M_M^2-s} \right).
\end{align}
Obtaining a smooth phase shift with correct threshold and asymptotic behaviour requires a description of the resonance beyond a simple Breit--Wigner parameterisation.
One should also account for the fact that the $\rho$ is actually not a stable particle, but can decay itself; one way is to include the self-energy of both the resonance $M$ and the $\rho$~\cite{Heuser:2024biq}.
One can go even further and relate these $\rho\pi$ phase shifts to three-body calculations for these resonances as in Refs.~\cite{Mai:2021nul,Sadasivan:2021emk,Yan:2025mdm}.

%---------------------------------------------------------------------------------------------------
\section{Determination of the TFFs}
\label{sec:Matching}
%---------------------------------------------------------------------------------------------------
An analogous result to \autoref{eq:PWsRhoPiRhoPi} can be obtained for resonance exchanges in the $\rho\pi \to \gamma^*\gamma^*$ amplitude by replacing one of the $\M_{M\rho\pi}^{\mu\tilde{\alpha}}$ in \autoref{eq:realpole_rhopirhopi} by the amplitude of $M$ coupling to two photons, 
\begin{align}
    \M_{M\gamma^*\gamma^*}^{\mu\nu\tilde{\alpha}} = \sum_i \tilde{T}_i^{\mu\nu\tilde{\alpha}} F_i^M(q_1^2,q_2^2).
\end{align}
These can be constructed via BTT~\cite{Hoferichter:2020lap} and define the TFFs $F_i^M(q_1^2, q_2^2)$.
In this way, we calculate partial-wave-projected helicity amplitudes $m_{\exch,j}^{M,J}(s)$ featuring the TFFs, which can be matched to the MO solution $m_{\MO,j}^{M,J}(s)$ containing the rescattering in $\rho\pi$ as well as $\rho$ and $\pi$ exchanges in the $t$- and $u$-channel via the Born terms.
This solution is constructed for each resonance $M$ from the usual Omnès functions, which only depend on the phase shifts, 
\begin{align}
    \Omega_M(s) = \exp \left( \frac{s}{\pi} \int_{(M_\rho+M_\pi)^2}^\infty \dff s^\prime \, \frac{\delta_M(s^\prime)}{s^\prime (s^\prime -s)}\right),
\end{align}
while the Born terms and the coupling between different helicity amplitudes and partial waves enter via the inhomogeneities
\begin{align}
     \Delta_j^J(s) = m_j^{\Born,J}(s) + \sum_{j^\prime, J^\prime} \frac{1}{2\pi\iu} \int_{(M_\rho+M_\pi)^2}^\infty \dff s^\prime \, K_{JJ^\prime}^{jj^\prime}(s, s^\prime) \, \disc_{s^\prime}^{\rho\pi} m_{j^\prime}^{J^\prime}(s^\prime).
\end{align}
If coupling between different partial waves is neglected and the kernel functions are diagonalised via a change of basis
\begin{align}
    m_j^J(s) \mapsto A_{jk}^J(s) m_k^J(s) =: \check{m}_j^J(s)
\end{align}
such that
\begin{align}
    \check{K}_{JJ^\prime}^{jj^\prime}(s, s^\prime) 
    = \frac{\delta_{JJ^\prime} \delta_{jj^\prime}}{s^\prime -s},
\end{align}
the system decouples to a set of standard MO problems with known solutions
\begin{align}
\check{m}_{\MO,j}^{M,J}(s) = \check{m}_j^{\Born,J}(s) + \frac{\Omega_M(s)}{\pi} \int_{(M_\rho+M_\pi)^2}^\infty \dff s^\prime \, \frac{\check{m}_j^{\Born,J}(s^\prime) \sin \delta_M(s^\prime)}{\vert \Omega_M(s^\prime) \vert (s^\prime -s)}.
\end{align}
Relating this to the original basis of partial-wave-projected helicity amplitudes and solving the matching equations
\begin{align}
    m_{\MO,j}^{M,J}(s) = m_{\exch,j}^{M,J}(s)
\end{align}
for the TFFs of $M \in \{a_1,a_2\}$ then yields a dispersive description of the latter, with input relying on the assumption that $\rho\pi$ is the main decay channel of $a_1$ and $a_2$ and the possibility to systematically include experimental or lattice data for the phase shifts.

This construction built on the general principles of analyticity, unitarity, and crossing symmetry will enable us to use available experimental, theoretical, and lattice input for constraining the TFFs of the isovector resonances.
The results then can be compared to the TFFs of the isoscalar partners $f_1$ and $f_2$. 
Especially a hierarchy between different TFFs of one intermediate state would be of interest for the HLbL contribution.

%---------------------------------------------------------------------------------------------------
\section*{Acknowledgements}
\label{sec:acknowledgements}
%---------------------------------------------------------------------------------------------------
The author thanks Leon Heuser, Martin Hoferichter, Bastian Kubis, Eirini Lymperiadou, Maxim Mai, Pablo Sánchez-Puertas, Peter Stoffer, and Maximilian Zillinger for helpful discussions and Martin Hoferichter and Bastian Kubis for comments on the manuscript.

\bibliographystyle{JHEP.bst}
\bibliography{bibliography} 

\end{document}